\documentclass[12pt,a4paper]{amsart}
\usepackage[utf8]{inputenc}
\usepackage[T1]{fontenc}
\usepackage{times,amsmath,amssymb,color,graphicx,graphics}
\usepackage{url,cite}

\date{\today}
\newcommand{\cn}{\,{\sf cn}}
\newcommand{\sn}{\,{\sf sn}}
\newcommand{\dn}{\,{\sf dn}}
\newcommand{\sech}{\,{\sf sech}}
\newcommand{\csh}{\,{\sf cosh}}
\newcommand{\tnh}{\,{\sf tanh}}
\newtheorem{theorem}{Theorem}
\newtheorem{remark}[theorem]{Remark}

\begin{document}
\title[Remarks on solutions to higher order KdV equations]{Remarks on existence/nonexistence of analytic solutions to higher order KdV equations}

\author{Anna Karczewska}
\address{Faculty of Mathematics, Computer Science and Econometrics, University of Zielona G\'ora, Szafrana 4a, 65-246 Zielona G\'ora, Poland.}
\email{A.Karczewska@wmie.uz.zgora.pl}

\author{Piotr Rozmej}
\address{Faculty of Physics and Astronomy, University of Zielona G\'ora, Szafrana 4a, 65-246 Zielona G\'ora, Poland}
\email{P.Rozmej@if.uz.zgora.pl}


\date{\today}

\begin{abstract}
In this note, we discuss the existence of analytic solutions to the nonlinear wave equations of the higher order than the ubiquitous Korteweg-de Vries (KdV) equation. First, we recall our recent results which show that the extended KdV equation (KdV2), that is, the equation obtained within second-order perturbation approach possesses three kinds of analytic solutions. These solutions have the same functional form as the corresponding KdV solutions. We show, however, that the most intriguing multi-soliton solutions, known for the KdV equation, do not exist for KdV2. Moreover, we show that for the equations obtained in the third order perturbation approach (and then in any higher order) analytic solutions in the forms known from KdV theory do not exist.
\end{abstract}

\maketitle


\section{Introduction} \label{intro}

The general problem of fluid motion with arbitrary boundary conditions leads to a set of \emph{Navier-Stokes equations}. In most cases attempts to solve these equations lead to extremely difficult problems. Therefore in many cases,  
some simplified models are introduced. For shallow water problem, physicists use the ideal fluid model. This means that fluid is assumed to be incompressible and inviscid with additional assumption that the fluid motion is irrotational. Since in normal conditions water viscosity and compressibility are very small the model should reproduce the fluid motion with reasonable accuracy, until waves on the surface do not break. So, for ideal fluid one obtains a system of four partial differential equations on two unknown functions, the velocity potential $\phi(x,y,z,t)$ and the surface elevation $\eta(x,y,t)$. This system contains the Laplace equation for velocity potential, kinematic and dynamic boundary conditions at the (unknown) surface and the kinematic boundary condition at the bottom. At this stage coordinates and time are dimensional quantities.

The introduction of scaled dimensionless coordinates to the system of the hydrodynamic Euler equations allows us to apply a perturbation approach. The solution of the velocity potential is assumed in the form of power series with respect to small parameter $\beta=(\frac{h}{l})^2$. Here, $h$ denotes the water depth, $l$ is an average wavelength. The boundary conditions at the hard bottom and the free surface introduce dependence on the second small parameter $\alpha=\frac{A}{h}$, where $A$ is the amplitude of the surface wave. 
The final wave equation for the surface wave can be obtained in different orders of the perturbation approach. It is worth noticing that the result of the perturbation approach depends on the relation between the small parameters. The case when $\alpha=O(\beta)$, that is, $\alpha$ and $\beta$ are of the same order, corresponds to weakly dispersive nonlinear waves and leads to the KdV equation \cite{KdV}. It is worth noticing that the KdV equation appears to be first order nonlinear wave equation in several different systems, like multilayer fluids, 
ion acoustic waves in plasma, electric circuits, and propagation of optical impulses in fibers, among others, see, e.g. monographs \cite{Whith,Ding,InfRow}.

In the fixed reference frame and scaled dimensionless coordinates the KdV equation has the following form  (indexes denote partial derivatives, e.g., $\eta_{3x}\equiv \frac{\partial^3 \eta}{\partial x^3}$)
\begin{equation} \label{nkdv}			
\eta_t+\eta_x+\frac{3}{2}\alpha\eta\eta_x + \frac{1}{6}\beta\eta_{3x}=0. 
\end{equation}

When the perturbation approach is continued to second order the resulted equation is known as the {\bf extended KdV} \cite{MS90} or {\bf KdV2}
\begin{equation} \label{nkdv2}	
\eta_t+\eta_x +\alpha\frac{3}{2} \eta\eta_x+\beta\frac{1}{6}\eta_{3x} -\alpha^2\frac{3}{8}\eta^2\eta_x + \alpha\beta \left(\frac{23}{24}\eta_x\eta_{2x} +\frac{5}{12}\eta\eta_{3x} \right) + \beta^2\frac{19}{360}\eta_{5x}=0.
\end{equation}

Taking into account all terms up to the third order in small parameters one arrives at the following {\bf KdV3} equation
\begin{align} \label{nkdv3}
\eta_t+\eta_x &+\alpha\frac{3}{2} \eta\eta_x+\beta\frac{1}{6}\eta_{3x} -\alpha^2\frac{3}{8}\eta^2\eta_x + \alpha\beta \left(\frac{23}{24}\eta_x\eta_{2x} +\frac{5}{12}\eta\eta_{3x} \right) + \beta^2\frac{19}{360}\eta_{5x} \nonumber\\ &
+\alpha^3\left(\frac{3}{16} \eta^3\eta_x\right) 
+ \alpha^2\beta \left(
\frac{19}{32}\eta_x^3+\frac{23}{16}\eta\eta_x\eta_{2x}+\frac{5}{16}\eta^2\eta_{3x}\right)  \\ &
 + \alpha\beta^2 \left(\frac{317}{288}\eta_{2x}\eta_{3x} +\frac{1079}{1440}\eta_{x}\eta_{4x} +\frac{19}{80}\eta\eta_{5x}  \right)+ \beta^3 \frac{55}{3024}\eta_{7x} =0. \nonumber 
\end{align}

The equations (\ref{nkdv})-(\ref{nkdv3}) are valid in the fixed reference frame. Mathematicians prefer simpler versions of KdV obtained by a transformation to a frame moving with a natural velocity (equal 1 in dimensionless coordinates or $\sqrt{gh}$ in dimension coordinates). For instance, in the case of  $\beta=\alpha$,  the following change of variables 
~$\hat{x} =\sqrt{\frac{3}{2}}\, (x-t), ~~\hat{t}= \frac{1}{4}\sqrt{\frac{3}{2}}\,\alpha t ~~\mbox{and} ~~u =\eta$~
transforms (\ref{nkdv}) into
$$
u_{\hat{t}} + 6 u u_{\hat{x}} + u_{3\hat{x}}=0
$$
commonly used in mathematical papers.

The paper is organized as follows. In section \ref{skdv} we give a short overview of analytic solutions to KdV. In section \ref{skdv2} the analytic solutions to KdV2, which have the same functional forms as the corresponding KdV solutions, are presented. We show that for KdV2 exact multi-soliton solutions do not exist. In section \ref{skdv3} we show that there are no analytic solutions for any  higher order KdV equations. 

\section{KdV and its solutions} \label{skdv}

The KdV equation possesses many miraculous properties. The most striking is the existence of an infinite number of integral invariants which correspond to conservation laws. There exist several kinds of analytic solutions to KdV:
single soliton solutions, periodic (cnoidal) solutions, periodic `superposition' solutions, and multi-soliton solutions. 
Two first kinds of these solutions can be obtained by direct integration (see, e.g., monographs \cite{Whith,Ding}). 
Below we show another method of obtaining single soliton and periodic solutions, applicable not only to KdV but also to KdV2. This method allows us to compare solutions of the same kind for equations of a different order.

\subsection{Single soliton solutions} \label{sssol}
Assume solutions of the KdV equation in the form
\begin{equation}\label{asumkdv}
\eta(x,t) = A \text{sech}^2[B(x-vt)]= A\,\text{sech}^2(By),
\end{equation}
where $y=x-vt$. Substitution of (\ref{asumkdv}) into KdV (see, equation (\ref{nkdv})) gives
\begin{equation}\label{akdv}
-\frac{1}{3} A B\tnh (B y) \sech^4(B y) [G_0+ G_2\csh (2 B y)]=0.
\end{equation}
Equation (\ref{akdv}) is valid for any argument only when simultaneously
\begin{equation}\label{G0}
G_0 =3 - 3 v+ 9\alpha A-10 \beta B^2 =0, \quad \mbox{and} \quad
G_2 =3 - 3 v +2 \beta  B^2 =0. 
\end{equation}
This gives immediately
\begin{equation}\label{Bv}
B=\sqrt{\frac{3\,\alpha}{4\,\beta} A}, \qquad v=1+\frac{\alpha}{2} A
\end{equation}
and the solution coincides with that obtained by direct integration of (\ref{nkdv}).

\begin{remark}  \label{rem1} 
It is clear from (\ref{Bv}) that solutions exist for arbitrary parameters $\,\alpha,\beta$. Since KdV imposes only two constraints on three coefficients $A,B,v$, there exists a one-parameter family of solutions. Usually, the amplitude $A$ can be considered arbitrary, within the range in which $\alpha$ stays small. 
\end{remark}

\subsection{Periodic (cnoidal) solutions} \label{ssper}
In this case, the solution is postulated in the form of the cnoidal wave
\begin{equation}\label{cnoid}
\eta(x,t) =A\,\cn^2[B(x-vt),m]+D.
\end{equation}
Equivalently, instead of the Jacobi elliptic $\cn$ function, $\dn$ or $\sn$ Jacobi elliptic functions can be used.
Then, substitution of (\ref{cnoid}) into KdV yields equation analogous to (\ref{akdv}), $
\frac{1}{3}AB\cn\sn\dn \left[G_0 +G_2 \cn^2 \right] =0.
$ 
So, there must be
\begin{equation} \label{g0}
G_0 = 4 \beta  B^2-8 \beta  B^2 m-9 \alpha D+6 v-6 =0,  \quad \mbox{and} \quad
G_2 = 12 \beta  B^2 m-9 \alpha  A= 0. 
\end{equation}
Equation $G_2=0$ implies 
\begin{equation}\label{B2pkdv}
 B =\sqrt{\frac{3\,\alpha}{4\,\beta}\frac{A}{m}}.
\end{equation}
Volume conservation condition determines
\begin{equation}\label{vccon1}
D = - \frac{A}{m} \left[\frac{E(m)}{K(m)}+m-1\right]. 
\end{equation}
In (\ref{vccon1}), $E(m)$ and $K(m)$ are the complete elliptic integral and  the complete elliptic integral of the first kind, respectively.  Then from $\,G_0=0$ one has
\begin{equation}\label{vpkdv1}
v=1 + \frac{\alpha A}{2 m}\left[2-m-3\,\frac{E(m)}{K(m)}\right].
\end{equation}

\begin{remark} \label{rem2}
In the case of cnoidal solutions, KdV with volume conservation condition supply three constraints on five parameters $A,B,v,D,m$. Then there is some freedom in allowable ranges of the coefficients. Usually, the amplitude $A$ is considered arbitrary, until there is no contradiction with the condition that $\alpha$ is small. In principle, the elliptic parameter can take values from the whole interval $m\in[0,1]$. 
\end{remark}

\subsection{Periodic `superposition' solutions} \label{sssup}

Assume solutions to KdV in the form 
\begin{equation} \label{eypm}
 \eta_{\pm}(y) = \frac{A}{2} \left[ \dn^2(By,m) \pm\sqrt{m}\cn(By,m)\dn(By,m) \right] +D.
\end{equation}
Coefficient $D$ is necessary in order to maintain, for arbitrary $m$, the same volume for a wave's elevations and depressions with respect to the undisturbed water level. This form of the solution (without $D$ term) has been proposed only recently in \cite{KhSa}.

Insertion of (\ref{eypm}) into (\ref{nkdv}) gives equation analogous to 
(\ref{akdv}), which after some simplifications takes the form
~$
F_0+F_2 \cn^2+ F_{11} \cn\dn   =  0. $
Then there are three conditions on the solution
\begin{align} \label{f0}  
F_0 & =9 \alpha  A-9 \alpha  A m-2 \beta  B^2+10 \beta  B^2 m+18 \alpha 
   D-12 v+12 =0, \\  \label{f2}
F_2 & =9 \alpha  A m-12 \beta  B^2 m  = 0, \\ \label{f11}
 F_{11} & = 9 \alpha  A \sqrt{m}-12 \beta  B^2 \sqrt{m} = 0.
\end{align}
Equations (\ref{f2}) and (\ref{f11})  are equivalent and yield the same 
\begin{align} \label{f2-11} 
B & = \sqrt{ \frac{3\alpha}{4\beta}\, A}.
\end{align}
Then volume conservation condition together with (\ref{f0}) determine $D$ and $v$ as functions of $A$ and $m$
\begin{equation} \label{d000} 
D= -\frac{A}{2} \frac{E(m)}{K(m)} \quad \mbox{and} \quad v = 1+ \frac{\alpha A}{8} \left[ 5-m-6\, \frac{E(m)}{K(m)}\right].
\end{equation}

The solutions $\eta_\pm$ (\ref{eypm}) are different than those given by (\ref{cnoid}). However, the remark \ref{rem2} applies to these solutions, as well.

\subsection{Multi-soliton solutions} \label{ssmult}

The existence of multi-soliton solutions is one of the most exciting properties  of the KdV equation. Zabusky and Kruskal \cite{ZabKru} noticed the first indication of that property in their famous numerical experiment. The paper \cite{ZabKru} inspired intensive studies which resulted in the development of a general method, by Gardner, Green, Kruskal and Miura \cite{Gardner}, called  \emph{IST (Inverse Scattering Transform)}. The IST allows us to construct the whole family of multi-soliton solutions. There exist also other methods for construction of multi-soliton solutions (e.g., B\"acklund transformations, Lax pairs, Hirota's direct method). During their motion solitons having different velocities collide and regain their shapes after reseparation. KdV equation permits for the existence of solitons with different amplitudes (and therefore different velocities). Therefore multi-soliton solutions exist.

\section{Analytic solutions to KdV2} \label{skdv2}

In this section, we discuss the solutions to KdV2 equation (\ref{nkdv2}) derived by us in the same way as solutions to KdV (\ref{nkdv}) in section \ref{skdv}. 
In \cite{KRI,IKRR,RKI,RK,KRbook}, we have shown that for the KdV2 equation there exist analytic solutions of the same forms as solutions to KdV (\ref{asumkdv}), (\ref{cnoid}) and (\ref{eypm}) but with different coefficients $A, B, D, v$. Here, we give a brief overview of these results, the full presentation of which is contained in \cite{KRI,IKRR,RKI,RK,KRbook}.

\subsection{Single soliton solution to KdV2} \label{sskdv2}

Substitution of the postulated form of the solution (\ref{asumkdv}) into (\ref{nkdv2}) leads (after some simplifications) to 
the equation analogous to (\ref{akdv}) but containing $C_0+C_2\sech^2(By)+C_4\sech^4(By)$ in the square bracket. This implies 
three linearly independent equations for three unknowns coefficients $A,B,v$
\begin{align} \label{C0}
(1-v) + \frac{2}{3}  B^2 \beta + \frac{38}{45}  B^4 \beta^2 &=0, \\ \label{C2}
 \frac{3\, A \alpha}{4} -  B^2 \beta + \frac{11}{4}A \alpha\, B^2  \beta -  \frac{19}{3} B^4 \beta^2 &=0,\\ \label{C4}
  -\left(\frac{1}{8}\right) (A \alpha)^2 - \frac{43}{12}A \alpha\, B^2  \beta +  \frac{19}{3}  B^4 \beta^2 &=0.
\end{align} 
From (\ref{C4}), denoting ~$ z=\frac{\beta B^2}{\alpha A}$~ one has 
\begin{equation} \label{rkw}
\frac{19}{3} z^2 - \frac{43}{12} z -\frac{1}{8}=0, 
\end{equation}
with roots 
\begin{equation} \label{rkw1}
 z_1 = \frac{43-\sqrt{2305}}{152}\approx -0.033  <0 \quad \mbox{and} \quad 
 z_2 = \frac{43+\sqrt{2305}}{152}\approx 0.599  > 0. 
\end{equation}
Since for soliton solution $A>0$, only ~$z=z_2$~ has physical relevance ($B^2>0$).
In this case (see \cite{KRI,KRbook}), all three coefficients of the solution are fixed by the coefficients of the KdV2 equation $\alpha,\beta$
\begin{equation} \label{ABv}
A \approx \frac{0.242399}{\alpha} >0,\quad 
B \approx \sqrt{\frac{0.145137}{\beta}},\quad v  \approx 1.11455.
\end{equation}

\subsection{Periodic cnoidal solutions to KdV2} \label{spkdv2} 

Substitution of the postulated form of the solution (\ref{cnoid}) into (\ref{nkdv2}) leads, similarly as in the soliton case discussed in subsection \ref{sskdv2}, to three linearly independent equations for four unknowns coefficients $A,B,D,v$ which have to be supplemented by volume conservation condition. The last one gives the relation $D=-\frac{A}{m}\left[ \frac{E(m)}{K(m)} + m-1\right]$, analogous to that in (\ref{d000}). Substitution $z = \frac{B^2\,\beta}{A\,\alpha} m$ transforms one of the equations 
into (\ref{rkw}). In this case, however, both positive and negative $z$ roots (\ref{rkw1}) can be relevant. For both $z$-roots the explicit formulas for $A,B,D,v$ are obtained as functions of $\alpha,\beta$ and the elliptic parameter $m$. Although from mathematical viewpoint all values of $m\in[0,1]$ are admissible, the defnition of small parameter $\alpha$ requires that the amplitude coefficient $A$ can not be much greater than one (in scaled coordinates).
This requirement puts limits on the physically relevant values of the elliptic parameter $m$. 

For the case $z=z_2$, physically reasonable values of $A$ occur only in a narrow interval of $m$ close to 1 (remember that when $m\to 1$ the space period of cnoidal wave tends to infinity and the solution tends to the soliton solution).
These solutions are `normal' cnoidal waves with the crests up and the troughs down, slightly different from cnoidal KdV solutions.

For the case $z=z_1$, however, the cnoidal KdV2 solutions exhibit a new feature.
$B$ is real for negative $A$. This fact means that the solutions are `inverted' cnoidal waves, with the crests down and the troughs up. The reasonable values of $|A|$ occur only for small $m\in [0, \approx 0.2)$. Therefore shapes of such waves are not much different from the usual cosine waves. 
For more details, we refer to \cite{IKRR,KRbook}.

\subsection{Periodic 'superposition' solutions to KdV2} \label{ssukdv2} 

In this case, the general procedure for obtaining the solutions of KdV2 equation is analogous to that for the KdV case, described in the subsection~\ref{sssup}.
Insertion of $\eta_\pm$ into the KdV2 equation (\ref{nkdv2}) supplies now the equation of the form $F_0 +F_2\cn^2+F_4\cn^4+F_{11}\cn\dn+F_{31}\cn^3\dn=0$. It appears, however, that only three of equations $F_i=0$ are linearly independent and with volume conservation condition constitute four equations for unknown coefficients of solutions. Substitution $z = \frac{B^2\,\beta}{A\,\alpha}$ transforms one of these equations into (\ref{rkw}). Then, similarly as in the case of cnoidal solutions, two cases occur, when $z=z_1$ and $z=z_2$. Finally, the `superposition' solutions to KdV2 are qualitatively similar to cnoidal solutions, but with slightly different amplitudes, velocities and wavelengths. For more details, we refer to \cite{RKI,RK,KRbook}.

\subsection{Nonexistence of multi-soliton solutions to KdV2} \label{snemkdv2} 
  
The given KdV2 equation (given values of $\alpha,\beta$ parameters) determines the unique single soliton solution (presented in subsection \ref{sskdv2}). This fact means that for  KdV2 there is no room for solitons moving with different velocities (and amplitudes). Therefore, {\bf multi-soliton solutions to KdV2 cannot exist}.

\section{Nonexistence of analytic solutions to higher order KdV equations} \label{skdv3}

Searching for single soliton solutions we
substitute the postulated form of the solution (\ref{asumkdv}) into (\ref{nkdv3}). This yields the equation
\begin{equation} \label{Kdv3}
C_0+ C_2\, \text{cosh}^2[B(x-vt)] +C_4\, \text{cosh}^4[B(x-vt)] + C_6\, \text{cosh}^6[B(x-vt)] = 0,
\end{equation}
where all $C_i$ are functions of $A,B,v$. The equation (\ref{Kdv3}) implies $C_0 = C_2 =C_4 =C_6 =0$. These equations are linerly independent and since the number of equations exceeds the number of unknowns they appear to be contradictory.
Therefore, single soliton solutions to KdV3 do not exist. 

Proceeding in the analogous way with search for cnoidal solutions (\ref{cnoid}) one reaches the similar structure of four conditions $C_0 = C_2 =C_4 =C_6 =0$. However, volume conservation condition supplies the fifth equation.
In this case, there are four unknowns $A,B,v$ and $D$. Again, similarly as in the previous case, the five equations for the unknown coefficients appear to be contradictory.

Qualitatively the same contradiction of equations is obtained for coefficients of `superposition' solutions to KdV3. In this case, the number of linerly independent equations exceeds the number of unknowns, as well.

{\bf Conclusion: For the KdV3 equation single soliton solutions (\ref{nkdv}), 
cnoidal solutions (\ref{cnoid}) and `superposition' solutions (\ref{eypm}) do not exist}.

\end{document}